# Audio-Based Music Classification with DenseNet And Data Augmentation

Wenhao Bian[1,2], Jie Wang[2], Bojin Zhuang[2], Jiankui Yang[1], Shaojun Wang[2] and Jing Xiao[2]

[1] Beijing University of Posts and Telecommnications, China
[2] Ping An Technology (Shenzhen) Co., Ltd, China
bianwenhao@bupt.edu.cn, photonicsjay@163.com,
zhuangbojin232@pingan.com.cn, yangjk@bupt.edu.cn,
wangshaojun851@pingan.com.cn, xiaojing661@pingan.com.cn

**Abstract.** In recent years, deep learning technique has received intense attention owing to its great success in image recognition. A tendency of adaption of deep learning in various information processing fields has formed, including music information retrieval (MIR). In this paper, we conduct a comprehensive study on music audio classification with improved convolutional neural networks (CNNs). To the best of our knowledge, this the first work to apply Densely Connected Convolutional Networks (DenseNet) to music audio tagging, which has been demonstrated to perform better than Residual neural network (ResNet). Additionally, two specific data augmentation approaches of time overlapping and pitch shifting have been proposed to address the deficiency of labelled data in the MIR. Moreover, an ensemble learning of stacking is employed based on SVM. We believe that the proposed combination of strong representation of DenseNet and data augmentation can be adapted to other audio processing tasks.

## 1 Introduction

With the rapid development of digital technology, the amount of online music accumulates so dramatically that structuring large-scale music is becoming a fundamental problem. Since 2000s, music information retrieval (MIR) has been widely studied for important applications including recommendation systems of music. As one of the main top-level descriptors (Chathuranga, 2013), music genre is a kind of label generally created by human experts and used for categorizing. However, it is impossible to label the gigabyte music manually. Thus, automatic music genre classification has been considered as a great challenge and valuable for MIR systems.

Most of music classification problems mainly consist of two modules. One is the preprocessing of raw audio data, the other is the design of classifier model. As a crucial part of the system, preprocessed data is the key to the final classification accura-cy. Generally, there are three main ways to preprocess raw audio: (1) acoustic features extraction (Auguin, 2013); (2) spectrograms transformation (Costa, 2011) (3) using raw audio (Dielman, 2011). Before the blossoming of deep learning, a common way is to extract specific acoustic features and aggregate them as input using various machine-learning algorithms (Ogihara, 2003). However, this method requires intense engineering effort and professional knowledge. With the fast development of deep learning, convolutional neural network (CNN) has received much success in image recognition and been tried in the MIR field (Nakashika, 2012). On the other hand, labelled music audio is really deficient in this area due to high cost professional tagging of experts.

In this paper, we exploit the advanced DenseNet as the building block of CNN architectures to boost performance of music audio classification, achieving higher accuracy than ResNet and baseline. In the part of data processing, grayscale spectrograms transformed from music raw audio are used for feature engineering (Dieleman, 2014). To address the shortage of labelled audio data, music-specific data augmentation is realized with the time overlapping and pitch shifting of spectrograms. All of the results verify that the methods we proposed achieve improvements over the state-of-the-art models on both of the FMA-small dataset (Defferrard, 2017) and GTZAN (Sturm, 2013).

This paper is structured as follows. In Section 2, a brief overview on related work is provided. We then describe the feature extraction in detail in Section 3. Section 4 discusses our methodologies, followed by the experimental results and some discussion. Finally, Section 6 provides conclusions and describes potential future work.

## 2  Related Work

As one of the high-level descriptors, music genre is always associated with harmonic, rhythm, pitch and other acoustic features (Aguiar, 2018). In this sense, physical properties of audio signal have been studied for music analysis. For instance, Mel Frequency Cepstral Coefficients (MFCCs) have been proven effective in the analysis of structures of music signals (Mubarak, 2006). Similar to other hand-crafted features, MFCCs is still a lossy representation. In order to fully utilize the information from the audio signal, raw audio has been directly used (Dielman, 2011). However, results show the use of raw data did not exhibit better performance than spectrograms in classification tasks. Spectrogram retains more information than MFCCs but with lower dimension than raw audio, which is more suitable for classification tasks (Wyse, 2017).

As a typical neural network of deep learning, CNN has been extensively applied in various image recognition tasks. Recently, CNN has been adapted for audio recognition tasks (Gwardys, 2014). In this kind of tasks, audio data was first converted to 2D spectrograms and then classified with CNN. For instance, Lee et al. (Lee, 2009) applied CNN to promote the classification accuracy of music genre and artist. And in (Choi, 2016), the usual CNN network with 2D convolutional layers obtained state-of-the-art performance at that time, which demonstrated the effectiveness of feature ex-traction of CNN for diverse music classification tasks. Contrary to visual images, however, 2D convolution of spectrograms along the frequency axis is not musically plausible to some extent. Lately, Dieleman et al. (Dieleman, 2014) introduced the network structure with '1D-CNN' to process spectrograms in music classification.

Recently, a number of sophisticated CNN models have emerged to improve the performance of image recognition drastically. For example, He et al. introduced ResNet (He, 2015) with skip connections enabling a very deep CNN to be effectively trained. And Huang et al. (Huang, 2017) proposed DenseNet which exploited feature reuse through dense connections instead of skip connection. Owing to the rapid advance of CNN in computer vision (CV), the ResNet architecture (Kim, 2018) was successfully applied for music auto tagging by processing raw audio directly. In this paper, we explore a more advanced CNN architecture, i.e. DenseNet, to process spectrogram instead of raw audio for music classification. To avoid over fit of network training, data augmentation is effectively realized with the time overlapping and pitch shifting of original spectrograms.

## 3 Data Processing

### 3.1 Input Length

Spectrogram represents spectra sequences varied along with the time axis. Spectrogram preparation is key to successfully applying CNN on music genres classification. In this way, the music audio tagging is reformulated as an image classification task (Schluter, 2013).

To generate grayscale spectrograms of music, Sound eXchange (SoX) package has been used. The spectrogram is with a fixed height of 128 pixels representing frequency per frame and varied widths dependent on 50 pixels per second of audio.

Additionally, the size of the spectrogram is also a hyper-parameter. If the full-scale spectrogram of an entire song is used, network size of CNN could be enlarged significantly. Following prior experiments in (Tokozume, 2017), time slice $T = 1.0 \sim 2.5s$ is applied in this work. Thus, each slice takes approximately 2.56s long segments. Finally, grayscale spectrograms with $128 \times 128$ dimension is input for CNN training. In the phase of evaluation, classification outputs of each slice of one song can be ensemble as a single song-level prediction.

### 3.2 Data Augmentation

Data Augmentation is a technique to avoid overfit of model training by increasing the volume of data. Based on the unique characteristics of music, two methods to augment the data are proposed. One is the time overlapping which is an effective way in the field of image processing. To increase the valid data size, window moving of audio signals generates extra data by setting the overlap of 50%. And the other way is pitch shifting. With a small change of pitch of a song, its classification still works. Thus, shifting the pitch of songs by a half of the tone is done with SoX. To increase the diversity of the data, extra data generated from two methods was mixed together, resulting in 3 times more data than the original one

**Fig. 1**(a) illustrates 50% overlap of spectrograms and (b) compares the difference between the origin and the pitch shifting spectrogram.

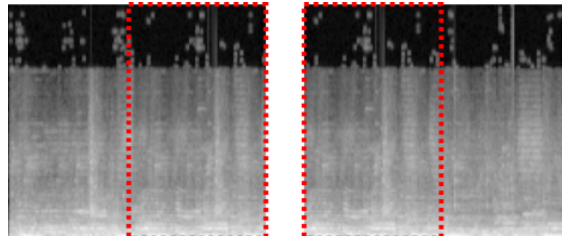

(a)  Time overlapping – 50%

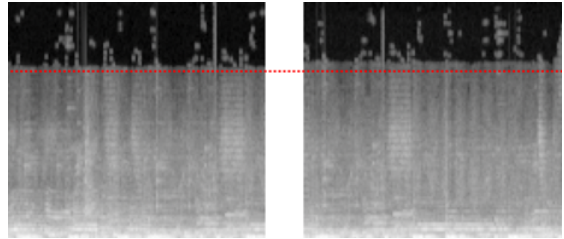

(b)  Pitch shifting – half tone higher

**Fig. 1.** Example of data augmentation techniques.

## 4   Methodology

Firstly, 1D CNN own less parameters and is computationally efficient (Nam, 2019), which is more suitable for music tagging because of limited dataset in this area. Moreover, 2D convolution over frequency dimension is uninterpretable (Ulyanov, 2016). Therefore, we choose 1D CNN as the basic block for spectrogram processing. Though time-consuming feature engineering can be alleviated by end-to-end training, the architecture of CNN network should be carefully designed for performance boost in specific tasks. In this paper, we firstly adapt DenseNet from CV to audio processing and compare its performance with ResNet and a regular CNN.

### 4.1   Basic Model

Our basic for music classification is shown in **Fig. 2(a)**, which is inspired by (Park, 2018). Gray-scale spectrograms with the size of 128×128 are prepared in the input layer. Instead of the popular 2D Conv, the convolution kernel of 1D Conv spans all the frequency of one time. As shown in **Fig. 2**, it mainly consists of 5 convolution layers with the kernel size of 4. After convolution layers, max-pooling layers with filter size of 4, 2 and 1 are applied in sequence. The last convolution layer with filter size of 1 is used instead of a fully connection layer (Lin, 2013). Batch normalization and rectified linear unit (RELU) are applied behind each convolution layer. Before the softmax output layer, there are two dense layers with 1024 and 8 hidden units respectively. The dashed boxes in **Fig. 2(a)** are the regular CNN blocks.

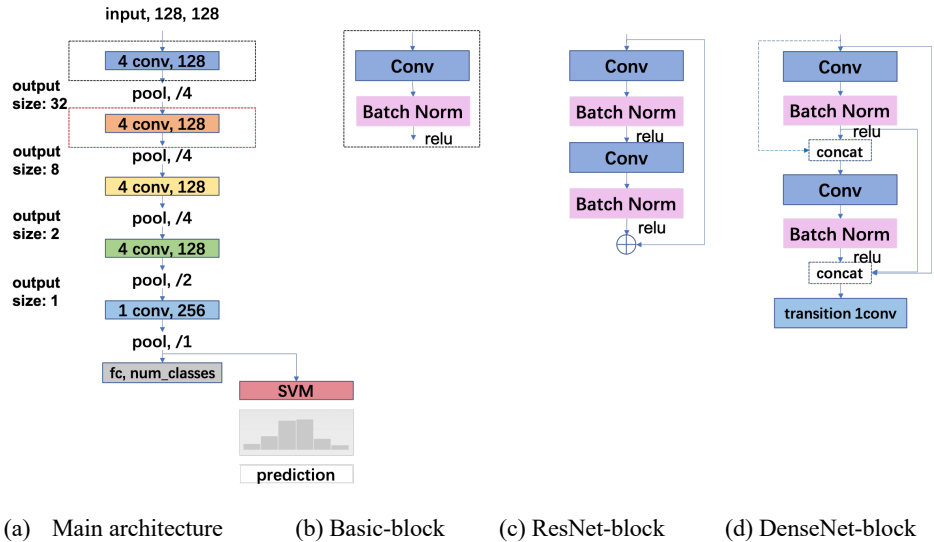

(a) Main architecture     (b) Basic-block     (c) ResNet-block     (d) DenseNet-block

**Fig. 2**. The proposed stacking model for music classification. (a) shows the main architecture with 5 convolution blocks and the SVM to predict the tags. (b) denotes the basic CNN block in the baseline model. (c)-(d) denotes the ResNet and DenseNet blocks to replace the basic one.

### 4.2 Advanced building blocks

**ResNet-block Model**. ResNet-block is inspired by residual learning (He, 2015). The basic block idea is that $H(x)$ is considered as the desired underlying mapping when $x$ is the input, then if one hypothesizes the net can approximate a residual function, i.e., $F(x) := H(x) - x$. The origin mapping becomes $F(x) + x$. Unlike (Kim, 2018), we modified the ResNet-block to adapt the spectrogram input rather than the raw audio. As shown in **Fig. 2**(c), the ResNet-block contains two convolution layers. After the block, the max pooling layer is applied. A comparison experiment has been conducted by replacing the basic block of the black dotted box in **Fig. 2(a)**.

**DenseNet-block Model**. The DenseNet-block is illustrated in **Fig. 2(d)**. In some sense, DenseNet (Huang, 2017) increases representational power of the basic CNN and capably improve performance over ResNet in classification tasks.

Contrary to the ResNet model, transition layers and connection methods are employed. Connectivity pattern in DenseNet is expressed $x_l = H_l([x_0, x_1, \ldots, x_{l-1}])$. the $L^{th}$ layer receives the feature-maps of all former layers that is different from the $H_l(x) = F(x_{l-1}) + x_{l-1}$, which can alleviate the gradient vanishing problem, and feature reuse can present better performance with fewer parameters and lower computational cost than ResNet. One dense-block contains two convolution layers with filter size of 4 and the translation layer follows the dense block to prevent the channel growing exponentially. In the same way, the DenseNet-block is used to replace the basic block of the red dotted box in **Fig. 2(a)**.

### 4.3 Ensemble of audio segments

During the data processing, a complete audio signal was clipped into small segments. Each segment is then predicted by the CNN classification model. Since these sliced segments belong to the same original audio clip, ensemble learning actually should be used. A common approach is to vote by selecting the most predicted labels among all segments as the final label. Inspired by (Gwardys, 2014), however, we propose to use SVM as the stacking classifier. Feature vectors can be obtained by averaging each feature vector of each segment extracted from the trained CNN model, and handled by SVM for final genre prediction.

## 5 Experiments and Results

### 5.1 Dataset

Our proposed architectures are experimented on two different datasets: FMA-small and GTZAN. FMA-small dataset is a new influential music dataset to help alleviate data scarcity problem of MIR (Defferrard, 2017), which contains 8000 tracks (.mp3 format) of 30 seconds per piece. There are 8 main genres with 1000 sub-classes per genre, such as Electronic, Experimental, Folk, Hip-hop, Instrumental, International, Pop, Rock. And most of them are with sampling rate of 44,100 Hz, bit rate 320 kb/s, and in stereo. The GTZAN dataset consists of 1000 audio tracks of 30 seconds long, which contains 10 genres with 100 tracks per genre. All tracks are 22,050Hz, Mono 16-bit audio files in .wav format (Sturm, 2013).

As described in Section 3, sound tracks are firstly processed into grayscale spectrograms and then is sliced into 10 segments with 2.56s per one. As a result, images of 128 ×128 are input the model with 128 frequency bins as the channel. Data augmentation described in Section 3 is also employed.

### 5.2 Training Details

In all experiments, model training is implemented with SGD and the learning rate of $1e^{-2}$ and the decay of $1e^{-6}$. Dropout of 0.5 is applied to the output of the full connection layer. Zero padding is applied to each convolution layer to maintain its size. In addition, mini batch of 128 samples and regularization of L2 are used to prevent the model from overfitting.

### 5.3 Results

**Table 1** summarizes the performance of baseline models with different parameters on FMA-small dataset. The kernel size of CNN cell is firstly optimized, where the model with kernel size of 4 outperform that of 3. In terms of ensemble methods, SVM offers higher accuracy than the voting method. When data augmentation is applied, significant improvement is achieved. Moreover, an interesting finding in **Table 1** shows that the ResNet-block in the red box performs better than that in the black box. In our opinion, 1D-CNN in the black box connect the entire frequency range at once, which is different from other convolution layers to some extent.

**Table 1.** Summarization of test accuracies of the Baseline and ResNet models on FMA-small dataset. Note that k4 denotes the kernel size of 4 while k3 is of 3, data-aug represents the data augmentation, ResNet–black denotes that the ResNet block is in the black box in Fig. 2(a), and ResNet-red means in the red box.

| Model + method | Accuracy (%) |
| --- | --- |
| Basic (k4 + voting) | 59.4 |
| Basic (k3 + SVM) | 61.3 |
| Basic (k4 + SVM) | 63.0 |
| Basic (k4 + SVM + data-aug) | 64.7 |
| ResNet-black (k4 + SVM + data-aug) | 63.7 |
| ResNet-red (k4 + SVM + data-aug) | 66.3 |

**Table 2.** Comparison with previous state-of-the-art models on FMA-small and GTZAN. Note that '~' denotes a round number because the accuracy is plot on the line chart but not provided.

| Model | FMA-small | GTZAN |
| --- | --- | --- |
| Transfer learning CNN (Gwardys, 2014) | - | 78.0 |
| MRMR (Baniya, 2014) | - | 87.9 |
| Transfer learning CNN (Choi, 2017) | - | 89.9 |
| SVM (Arabi, 2009) | - | 90.8 |
| Sparse representation-based classifier (Panagakis, 2010) | - | 93.7 |
| SVM (Defferrard, 2017) | 54.8 | - |
| Transfer learning CNN (Park, 2018) | 56.8 | - |
| Transfer learning CNN (Lee, 2019) | 51.2 (~) | 92.2 (~) |
| Our Basic Model | 64.7 | 84.0 |
| Our ResNet Model | 66.7 | 89.5 |
| Our DenseNet Model | 68.9 | 90.2 |

**Table 2** summarizes the results of our proposed DenseNet model compared to other state-of-the-art methods. On the FMA-small dataset, our models outperform all the previous results. Moreover, our proposed DenseNet model performs better than ResNet one owing to its strong feature extraction capability and cost less time to get a better result. Since the FMA-small dataset is published in 2017 and related experiments are not many, we adapted our method to another dataset, i.e. GTZAN to compare with more approaches. Obviously, that also shows the better performance.

As can be seen, even without extra dataset, our DenseNet model performs close to the state-of-the-art deep learning method by using transfer learning CNN (Lee, 2019). Note that our models which introducing the shortcut connections outperform (Park, 2018) which using the same basic model by enhancing the representation of a layer.

## 6      Conclusion and Future Work

In this paper, a comprehensive study on music classification using a DenseNet deep learning method is conducted. To overcome the shortage of labelled music audio data, a music-specific data augmentation method is proposed with time overlapping and pitch shifting on spectrograms. Owing to the strong feature extraction capability of DenseNet, our stacking method achieves a state-of-the-art result on FMA-small dataset. We believe our presented approach can be adapted to other audio processing tasks.

In the future, we will investigate our DenseNet Model more thoroughly. We plan to investigate every layer expression and then try diverse connections. Then, we will adapt more advanced architectures which are designed for the image and language challenges in audio tasks.

## 7      Acknowledgement

This work was supported by Ping An Technology (Shenzhen) Co., Ltd, China.